\documentclass[a4paper,11pt]{article}
\pdfoutput=1 

\usepackage{jinstpub} 

\title{A custom readout electronics for the BESIII CGEM detector}

\author[f,1]{M. Da Rocha Rolo\note{Corresponding author.}}

\author[f,l]{M. Alexeev}
\author[f,l]{A. Amoroso}

\author[a,c]{R. Baldini Ferroli}
\author[c]{M. Bertani}
\author[b]{D. Bettoni} 
\author[f,l]{F. Bianchi}
\author[h]{R. Bugalho}

\author[c]{A. Calcaterra}
\author[b]{N. Canale}
\author[c,e]{M. Capodiferro}
\author[b]{V. Carassiti}
\author[c]{S. Cerioni}
\author[a,f,i]{JY. Chai}
\author[b]{S. Chiozzi}
\author[b]{G. Cibinetto}
\author[f,i]{F. Cossio}
\author[b]{A. Cotta Ramusino}

\author[f,l]{F. De Mori}
\author[f,l]{M. Destefanis}
\author[h]{A. Di Francesco}
\author[c]{J. Dong}

\author[b]{F. Evangelisti}

\author[b,j]{R. Farinelli}
\author[f]{L. Fava}
\author[c]{G. Felici}
\author[b]{E. Fioravanti}

\author[b,j]{I. Garzia}
\author[c]{M. Gatta}
\author[f,l]{M. Greco}

\author[a,f]{L. Lavezzi}
\author[a,f,i]{CY. Leng}
\author[a,f]{H. Li}

\author[f,l]{M. Maggiora}
\author[b]{R. Malaguti}
\author[f,l]{S. Marcello}
\author[m]{P. Marciniewski}
\author[b]{M. Melchiorri}
\author[b,j]{G. Mezzadri}
\author[f]{M. Mignone}
\author[c]{G. Morello}

\author[d,k]{S. Pacetti}
\author[c]{P. Patteri}
\author[f,l]{J. Pellegrino}
\author[c,e]{A. Pelosi}

\author[f]{A. Rivetti}

\author[b,j]{M. Savri\'e}
\author[b,j]{M. Scodeggio}
\author[c]{E. Soldani}
\author[f,l]{S. Sosio}
\author[f,l]{S. Spataro}

\author[c,g]{E. Tskhadadze}

\author[h]{J. Varela}
\author[j]{S. Verma}

\author[f]{R. Wheadon}

\author[f]{L. Yan}

\emailAdd{darochar@to.infn.it}

\affiliation[a]{Institute of High Energy Physics, Chinese Academy of Sciences, 19B YuquanLu, Beijing, 100049, China}
\affiliation[b]{INFN, Sezione di Ferrara, via G. Saragat 1, 44122 Ferrara, Italy}
\affiliation[c]{INFN, Laboratori Nazionali di Frascati, via E. Fermi 40, 00044 Frascati (Roma), Italy}
\affiliation[d]{INFN, Sezione di Perugia, via A. Pascoli 14, 06123 Perugia, Italy}
\affiliation[e]{INFN, Sezione di Roma, c/o Universit\`a La Sapienza, p.le A. Moro 2, 00185 Roma, Italy}
\affiliation[f]{INFN, Sezione di Torino, via P. Giuria 1, 10125 Torino, Italy}
\affiliation[g]{Joint Institute for Nuclear Research (JINR), Joliot-Curie 6, Dubna, Moscow region, 141980, Russia}
\affiliation[h]{Laboratório de Instrumentação e Física Experimental de Partículas (LIP), Av Elias Garcia 14, 1000-149, Portugal}
\affiliation[i]{Politecnico di Torino, Dipartimento di Elettronica e Telecomunicazioni, Corso Duca degli Abruzzi 24, 10129 Torino, Italy}
\affiliation[j]{Universit\`a di Ferrara, Dipartimento di Fisica, via G. Saragat 1, 44122 Ferrara, Italy}
\affiliation[k]{Universit\`a di Perugia, Dipartimento di Fisica e Geologia, via A. Pascoli 14, 06123 Perugia, Italy}
\affiliation[l]{Universit\`a di Torino, Dipartimento di Fisica, via P. Giuria 1, 10125 Torino, Italy}
\affiliation[m]{Uppsala Universitet, Department of Physics and Astronomy, Lägerhyddsvägen 1, 752 37 Uppsala, Sweden}

\abstract{
For the upgrade of the inner tracker of the BESIII spectrometer, planned for 2018, a lightweight tracker based on an innovative Cylindrical Gas Electron Multiplier (CGEM) detector is now under development. 
The analogue readout of the CGEM enables the use of a charge centroid algorithm to improve the spatial resolution to better than $130\ \mu$m while loosening the pitch strip to $650\ \mu$m, which allows to reduce the total number of channels to about 10 000.
The channels are readout by 160 dedicated integrated 64-channel front-end ASICs, providing a time and charge measurement and featuring a fully-digital output. 

The energy measurement is extracted either from the time-over-threshold (ToT) or the 10-bit digitisation of the peak amplitude of the signal.
The time of the event is generated by quad-buffered low-power TDCs, allowing for rates in excess of 60 kHz per channel. 
The TDCs are based on analogue interpolation techniques and produce a time stamp (or two, if working in ToT mode) of the event with a time resolution better than 50 ps. 
The front-end noise, based on a CSA and a two-stage complex conjugated pole shapers, dominate the channel intrinsic time jitter, which is less than 5 ns r.m.s.. 
The time information of the hit can be used to reconstruct the track path, operating the detector as a small TPC and hence improving the position resolution when the distribution of the cloud, due to large incident angle or magnetic field, is very broad.

Event data is collected by an off-detector motherboard, where each GEM-ROC readout card handles 4 ASIC carrier FEBs (512 channels). 
Configuration upload and data readout between the off-detector electronics and the VME-based data collector cards are managed by bi-directional fibre optical links.

This paper covers the design of a custom front-end electronics for the readout of the new inner tracker of the BESIII experiment, addressing the relevant design aspects of the detector electronics and the front-end ASIC for the CGEM readout, and reviewing the first silicon results of the chip prototype.

}


\proceeding{INSTR17: Instrumentation for Colliding Beam Physics\\
  27 February - 3 March, 2017\\
  Novosibirsk, Russia}

\begin{document}
\maketitle
\flushbottom

\section{Introduction}

The Beijing Spectrometer III (BESIII) operates at the $\tau$-charm energy region and runs since 2009 at the $e^+$ $e^-$ collider (BEPCII), hosted by the Institute of High Energy Physics (IHEP) in Beijing \cite{a}.
The high luminosity of this multi-bunch collider, which has reached $1.0 \times 10^{33}\ cm^{-2}s^{-1}$ in 2016, increased the radiation dose on the inner Main Drift Chamber (MDC) tracker.
An upgrade program has been set up to counter this ageing effect and allow the operation of the BESIII experiment beyond 2022.
This upgrade proposal includes the development of a lightweight Cylindrical Gas Electron Multiplier Inner Tracker (CGEM-IT), and its installation is planned for mid 2018.

\begin{figure}[htbp]
\centering
\includegraphics[width=.27\textwidth]{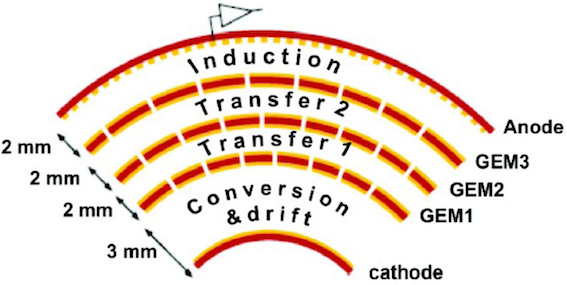}
\qquad
\includegraphics[width=.43\textwidth]{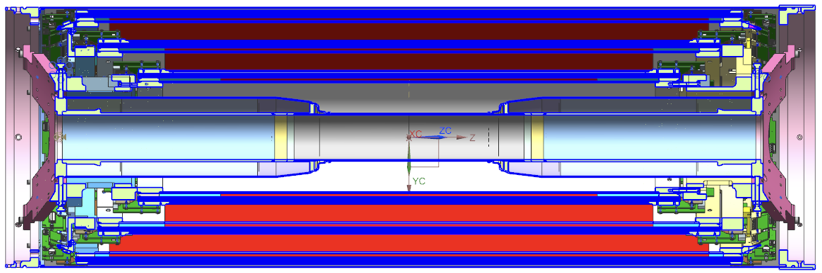}
\qquad
\includegraphics[width=.18\textwidth]{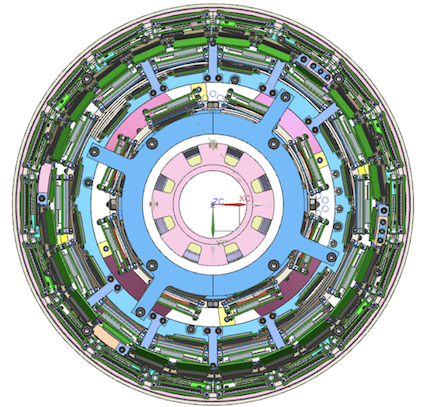}
\caption{\label{fig:cgem} Schematic of a Triple-GEM (\textit{left}) and 2D section views (\textit{centre, right}) of the CGEM detector mechanical drawings.}
\end{figure}

A Triple-CGEM (Figure \ref{fig:cgem}) builds up three concentrical independent tracking layers, each of one assembling a cathode, three GEM foils and the readout anode.
This design allows for higher gain with lower discharge rates in respect to the single GEM, and covers 93\% of the solid angle. 
The readout anode of each CGEM is segmented with XV patterned strips with a 650 $\mu$m pitch.
Complementary charge centroid and $\mu$-TPC algorithms are then used to provide an $x-y$ position resolution of 130 $\mu$m \cite{b} on the track reconstruction.
The analogue readout of the detector, which allows reducing the total number of channels to around 10 000, requires dedicated front-end electronics.
This work describes and discusses the design, characterisation and production plans for the on-detector integrated and front-end circuitry, off-detector data collection and power electronics.

\section{Overview of the CGEM Readout Electronics}

\begin{figure}[htbp]
\centering
\includegraphics[width=.4\textwidth]{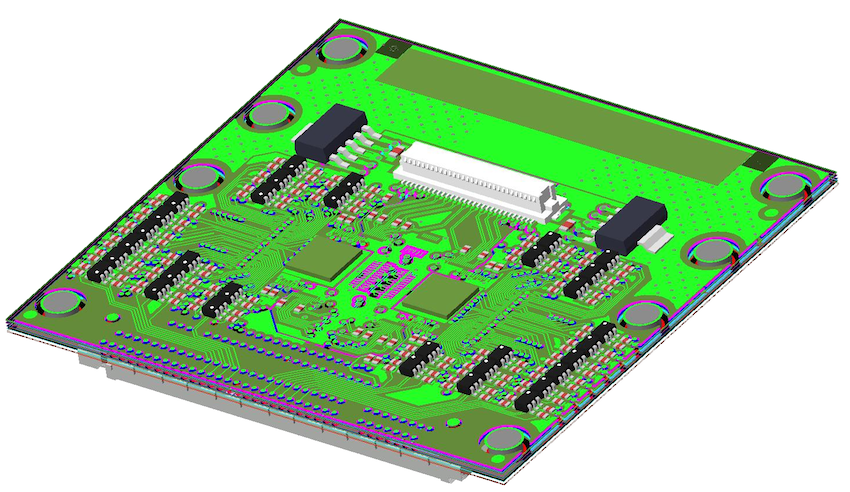}
\qquad
\includegraphics[width=.4\textwidth
]{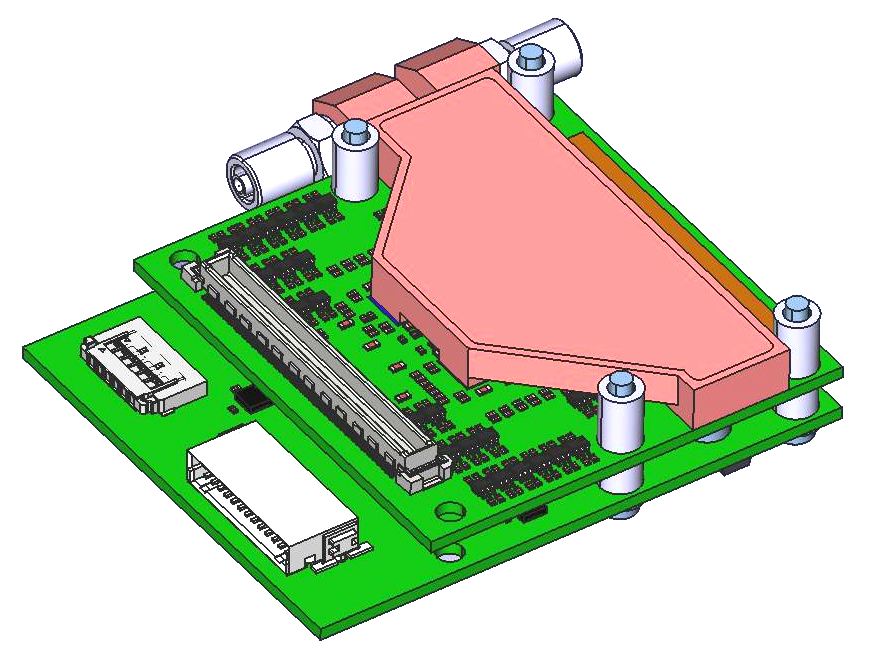}
\caption{\label{fig:feb} Front-End Board Design for Layer 1: top-side routing on FE1 (\textit{left}) and FE1/FE2 assembly with liquid cooling heat exchanger plate (\textit{right}).}
\end{figure}

The analogue readout of the Triple-GEMs employs a dedicated multi-channel front-end Application-Specific Integrated Circuit (ASIC), developed specifically to target the design requirements of the BESIII-CGEM detector.
Two 64-channel chips are mounted on each one of the 80 Front-End Boards (FEBs). 
The FEB (Figure \ref{fig:feb}) consists on a stack of an analogue-most layer (FE1), hosting 2 ASICs, regulators and ESD protection networks, interface towards the anode and the digital-domain FE2, which handles the on-detector electronics signal, data and power interface.

\begin{figure}[htbp]
\centering
\includegraphics[width=.95\textwidth]{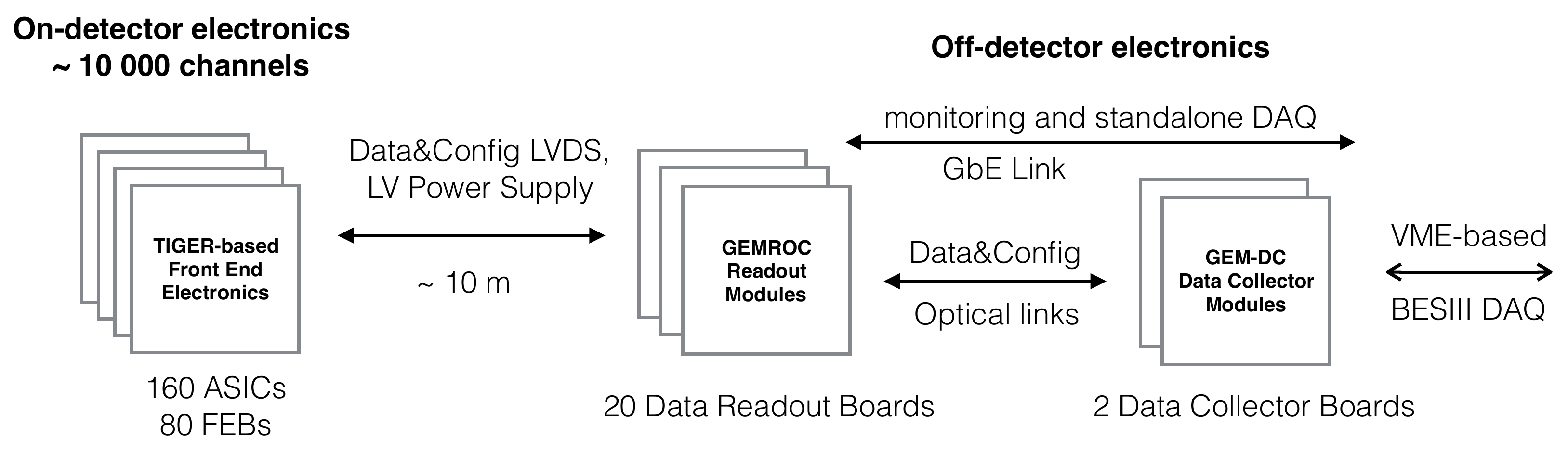}
\caption{\label{fig:electronics} Conceptual block diagram overview of the CGEM Detector Electronics}
\end{figure}

Off-detector GEM Readout Cards (GEMROC) modules handle quad-FEBs through 10-metre LVDS links for data, configuration and monitoring.
The GEMROC uses an ALTERA ArriaVGX FPGA development board coupled, through a High-Speed Mezzanine Card (HSMC) high performance connector, to the Interface Card (GEMROC\_IFC), which manages the electrical and physical interfaces to the FEBs, to the GEM-DC data concentrator (bi-directional fibre optic links) and the BES-III Fast Control system. 
An ethernet port is also available for monitoring and debugging.
Each GEM-DC data collector is based on the Advanced Trigger Logic Board (ATLB) \cite{e} and handles up to 16 optical links running at 2 Gbit/s (max 6.5 Gb/s).
It features a VME base board for the interface with the BESIII Data Acquisition (DAQ) system.
Both the GEMROCs and the on-detector electronics are powered by a dedicated Low-Voltage (LV) distribution system, allowing for single board current/voltage monitoring, fuses and on/off capability, and remote controlled via ethernet.
Figure \ref{fig:electronics} illustrates the block diagram of the detector electronics.

\section{The TIGER ASIC}

\begin{figure}[htbp]
\centering
\includegraphics[width=.4\textwidth]{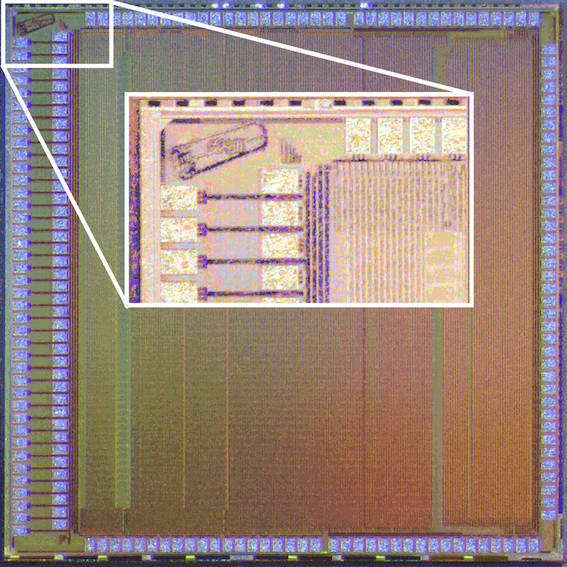}
\caption{\label{fig:tiger} Microphotograph of the TIGER chip.}
\end{figure}

The Torino Integrated GEM Electronics for Readout (TIGER) chip is a 64-channel mixed-mode circuit for the readout of GEM detectors.
The $5 \times 5\ mm^2$ ASIC is designed on an UMC CMOS 110nm technology node (Figure \ref{fig:tiger}).

\subsection{Design of a mixed-mode 64-channel ASIC for Micropattern gaseous detectors }

The ASIC features a linear array of 64 channels, which provide amplification, shaping and signal conditioning, time and charge measurement, at a maximum data rate of 100 kHz. 
The time stamp and digitisation data are collected by a back-end controller, upgraded from IP used in a family of chips designed for medical imaging \cite{c, d}.
This upgrade adds single-event-upset protection (SEU) for finite-state machines (with TMR) and configuration payload (Hamming encoding).
The 8B/10B encoded fully digital data output uses up to 4 LVDS Tx drivers operating in single or double-data-rate at maximum rate of 640 Mb/s per link.
An SPI protocol is used for the configuration upload/download, where a set of periphery DACs is used for the biasing of the channel analogue circuitry.
On-chip calibration circuits generate programmable test charge pulses to each channel, used for calibration (before and after installation) and debug purposes.

\begin{figure}[htbp]
\centering
\includegraphics[width=.8\textwidth]{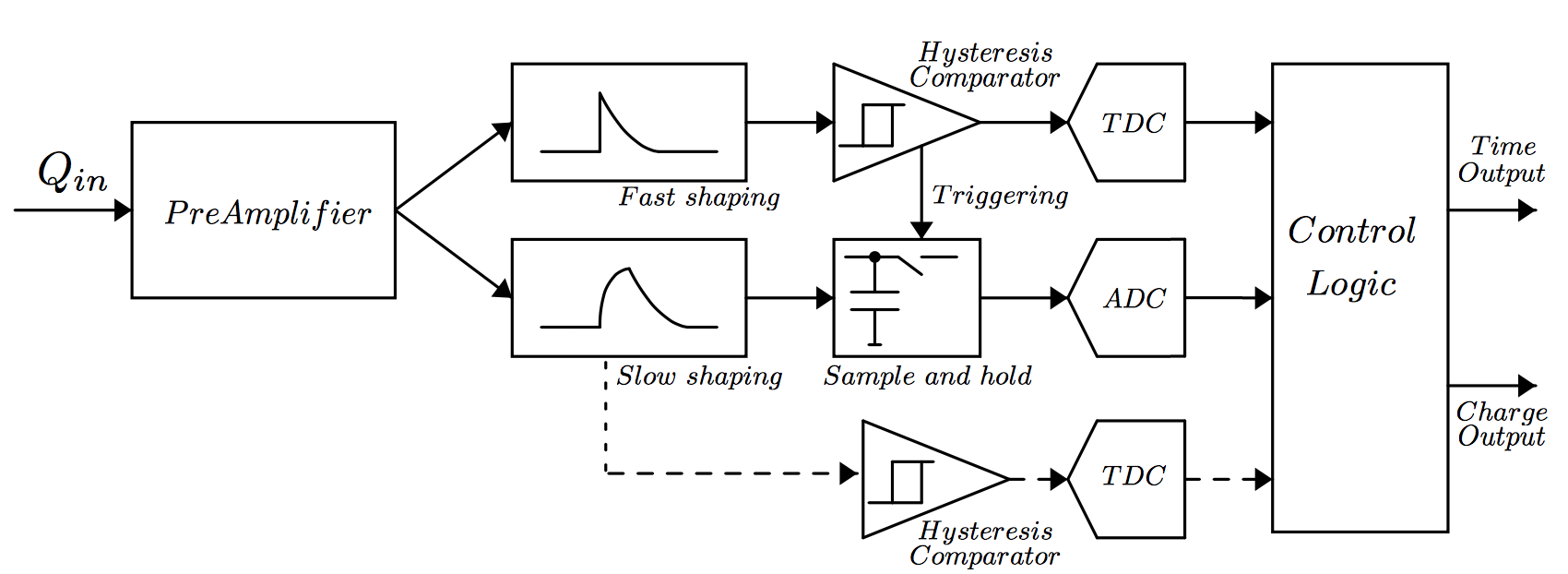}
\caption{\label{fig:channel} Block diagram of the TIGER channel.}
\end{figure}

Figure \ref{fig:channel} illustrates the block diagram organisation of the TIGER channel.
A time-based readout operation mode, using single or double threshold, produces a time stamp on the rising and falling edges of the signals produced by the fast and slow shaping, respectively.
The timing measurement is obtained with a sub-50 ps binning (with a system clock of 160 MHz) quad-buffered time-to-digital converters (TDCs) based on time interpolators \cite{c}. 
When working in a time-based readout, the charge information can be retrieved from the Time-over-Threshold (ToT).
The inherent non-linear measurement requires a ToT vs. Q$_{in}$ calibration, which compression curve is typical when the time-over-threshold is measured with CR-RC$^2$ shapers.

\begin{figure}[htbp]
\centering
\includegraphics[width=.45\textwidth]{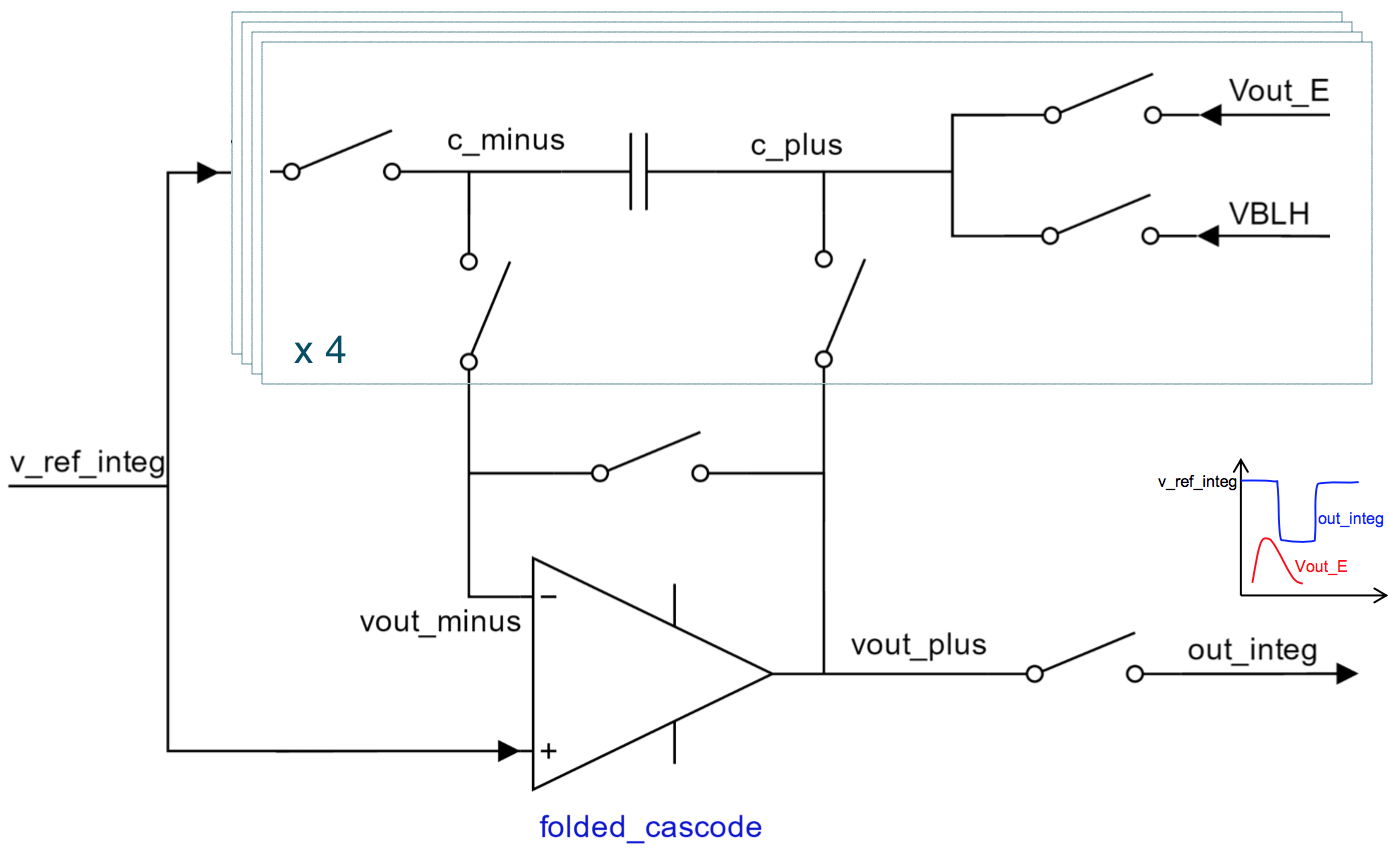}
\qquad
\includegraphics[width=.45\textwidth]{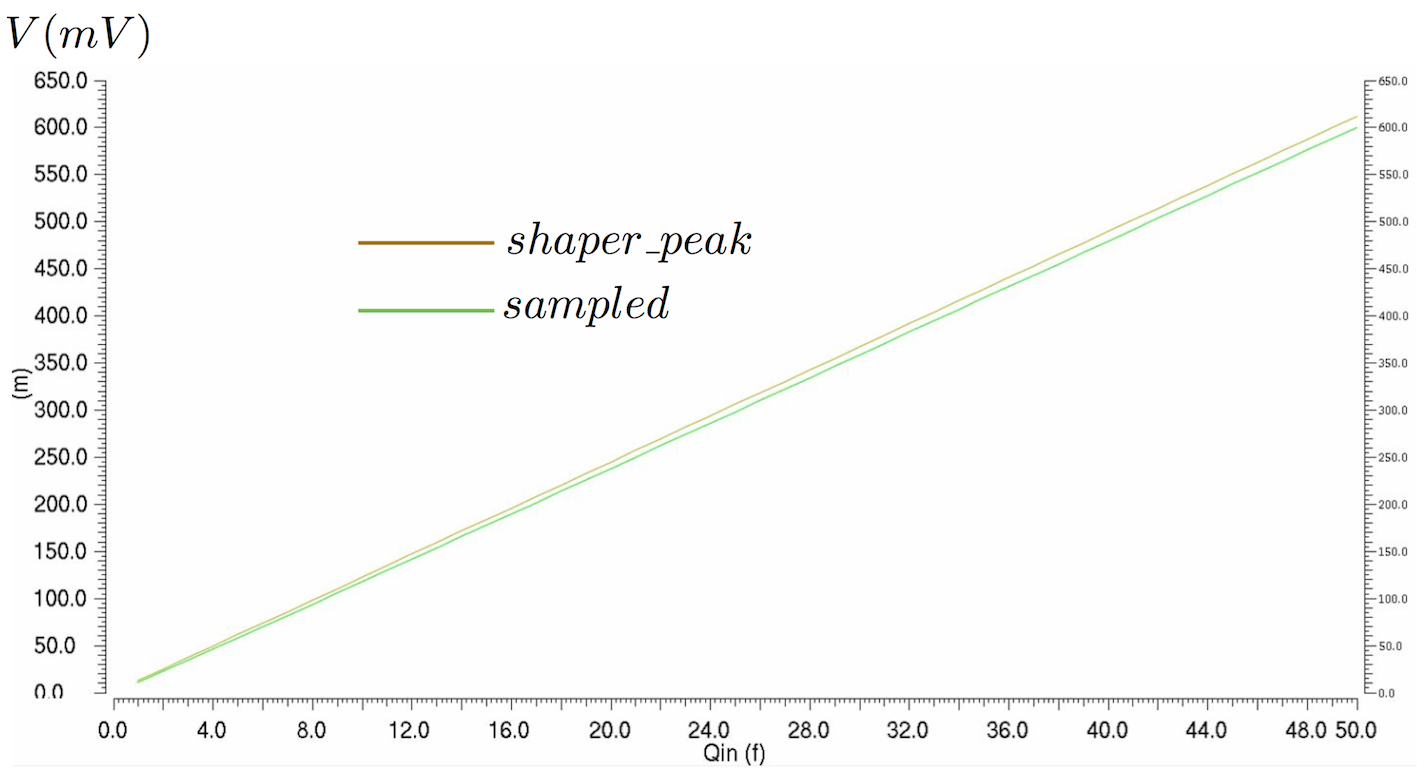}
\caption{\label{fig:sh} Sample-and-Hold circuit for amplitude measurement; circuit implementation (\emph{left}) and expected linear behaviour on the measurement of the signal peak and sampled amplitude as a function of the input charge (\emph{right}).}
\end{figure}

In standard operation with GEMs, a Sample-and-Hold circuit will be used instead for the energy measurement (Figure \ref{fig:sh}).
The circuit consists on a programmable digitally-controlled peak detector, where the start of the sampling is set by the trigger on the fast discriminator output, and the sampling time can be set in steps of 25 ns.
The digitisation of the voltage amplitude is performed with a 10-bit Wilkinson ADC, otherwise used for the fine time measurement of the slow shaper falling edge trigger.
While this simpler strategy, in respect to a two-phase analogue peak detector, avoids the use of rail-to-rail input operational amplifiers, it is inherently more susceptible to the jitter and time-walk on the leading-edge discriminators.
With the nominal expected jitter from simulations, we expect an error on the sampled voltage smaller than 1\%, which is adequate for the purposes of the CGEM detector.

The very-front-end is optimised to the readout of signals in the range 1 - 50 fC, and targets a noise below 2000 electrons r.m.s. for a channel detector capacitance of 100 pF.
A low-noise two-stage cascode charge sensitive front-end generates two replicas of the amplified signal, which is split for time and charge measurement branches in two dedicated shapers.
The two shapers produce a semi-Gaussian output signal shape with a peaking time 60 and 170 ns, respectively for the time and charge branches.
The maximum signal width expected is around 1 $\mu s$ on the energy shaper, thereby limiting the pile-up probability to less than 1\% at 60 kHz.
A baseline-holder circuit locks the DC voltage at the output of the amplifier to a voltage that can be set externally between 250 and 650 mV, typically set to 300 mV.

\subsection{Electrical Characterisation Results}

\begin{figure}[htbp]
\centering
\includegraphics[width=.4\textwidth]{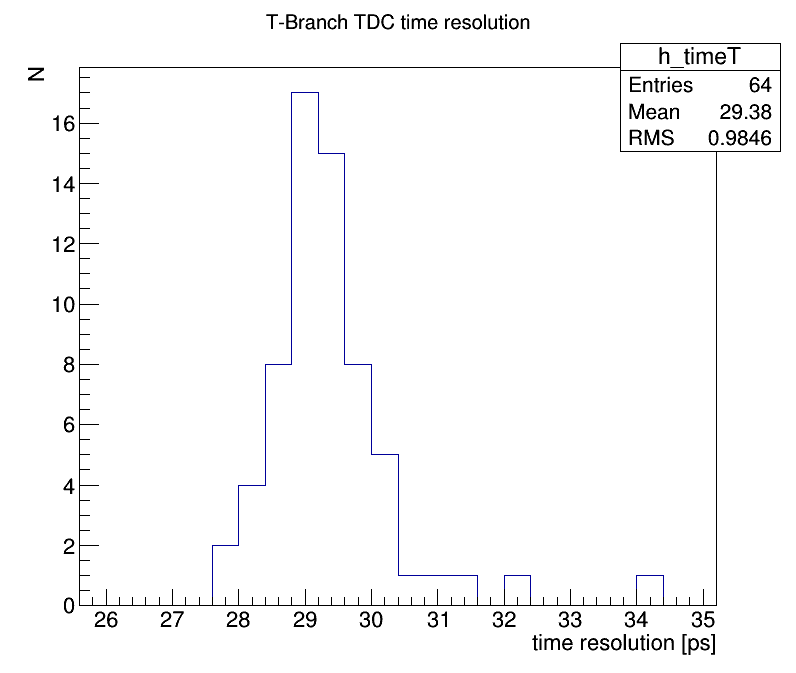}
\qquad
\includegraphics[width=.5\textwidth]{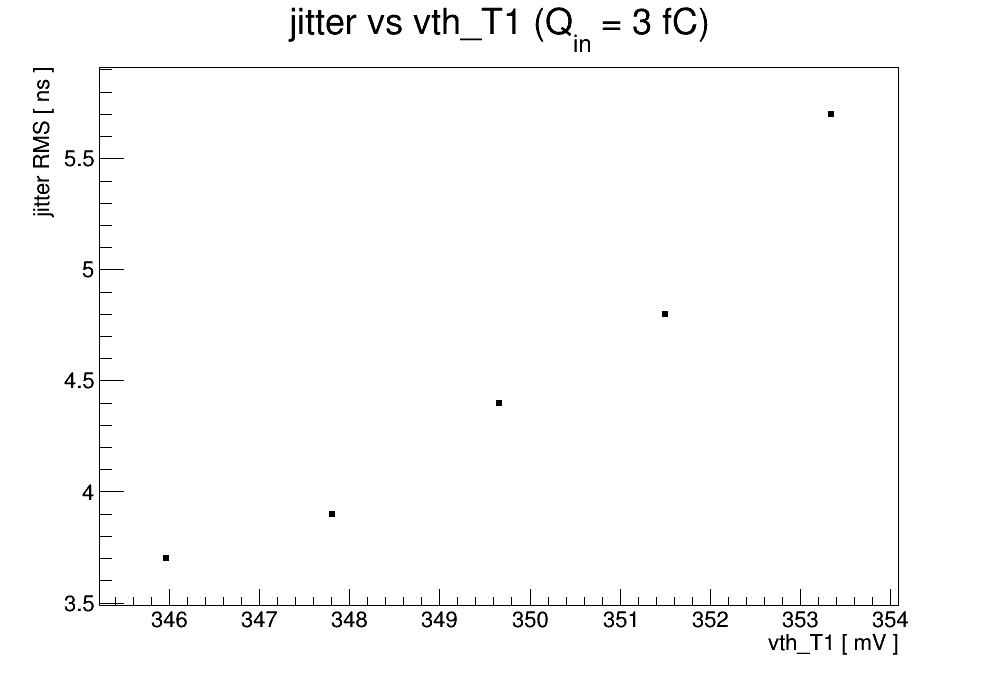}
\caption{\label{fig:time} Time resolution of the TDC on the 64 channels (\emph{left}); and jitter on the time measurement for $C_{d} = 100\ pF$ and $Q_{in} = 3\ fC$ as a function of the threshold (\emph{right}).}
\end{figure}

The first TIGER prototype was produced with a Multi-Project-Wafer (MPW) shuttle and the electrical characterisation campaign started in October 2016.
Read/Write operations of channel/global configuration registers, data transmission and decoding, (dual-) TDC operation and fine calibration and the sample-and-hold circuit work as expected.
Figure \ref{fig:time} shows the time resolution of the TDC on 64 channels after calibration.
Such calibration, performed with a sweep of the test-pulse along one clock cycle, generates a look-up table (LUT) that stores the values of the interpolation factor and offset for each TDC/interpolator.
The average 30 ps r.m.s. quantisation error is much lower than the required jitter on the measurement of the event, and the intrinsic time resolution of the channel is only affected by the noise of the front-end.
For an input capacitance of 100 pF and a signal level or 3 fC, the jitter on the time measurement is typically below 5 ns r.m.s. (refer to figure \ref{fig:time}, right plot).

The noise of the front-end is estimated after a threshold scan with a fixed input charge generated by the on-chip calibration circuit.
A sigmoid fit to the resulting s-curve provides a r.m.s. value for the intrinsic noise, and this procedure is repeated for different input capacitance values or configuration and operation mode settings.
The same method is used for the baseline scan, which results are saved into a per-channel threshold LUT that allows for the equalisation of the effective threshold.

\begin{figure}[htbp]
\centering
\includegraphics[width=.46\textwidth]{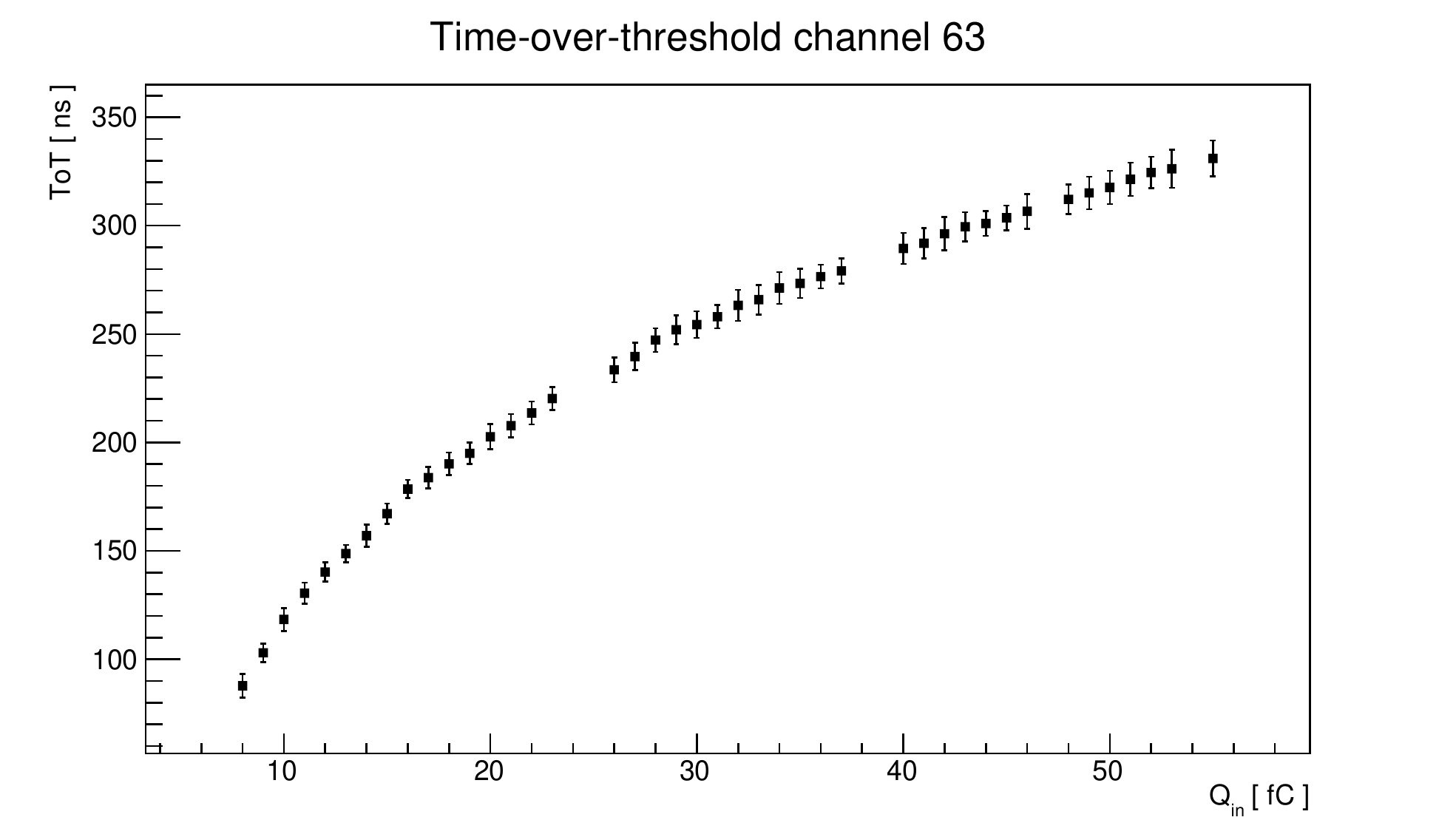}
\qquad
\includegraphics[width=.42\textwidth]{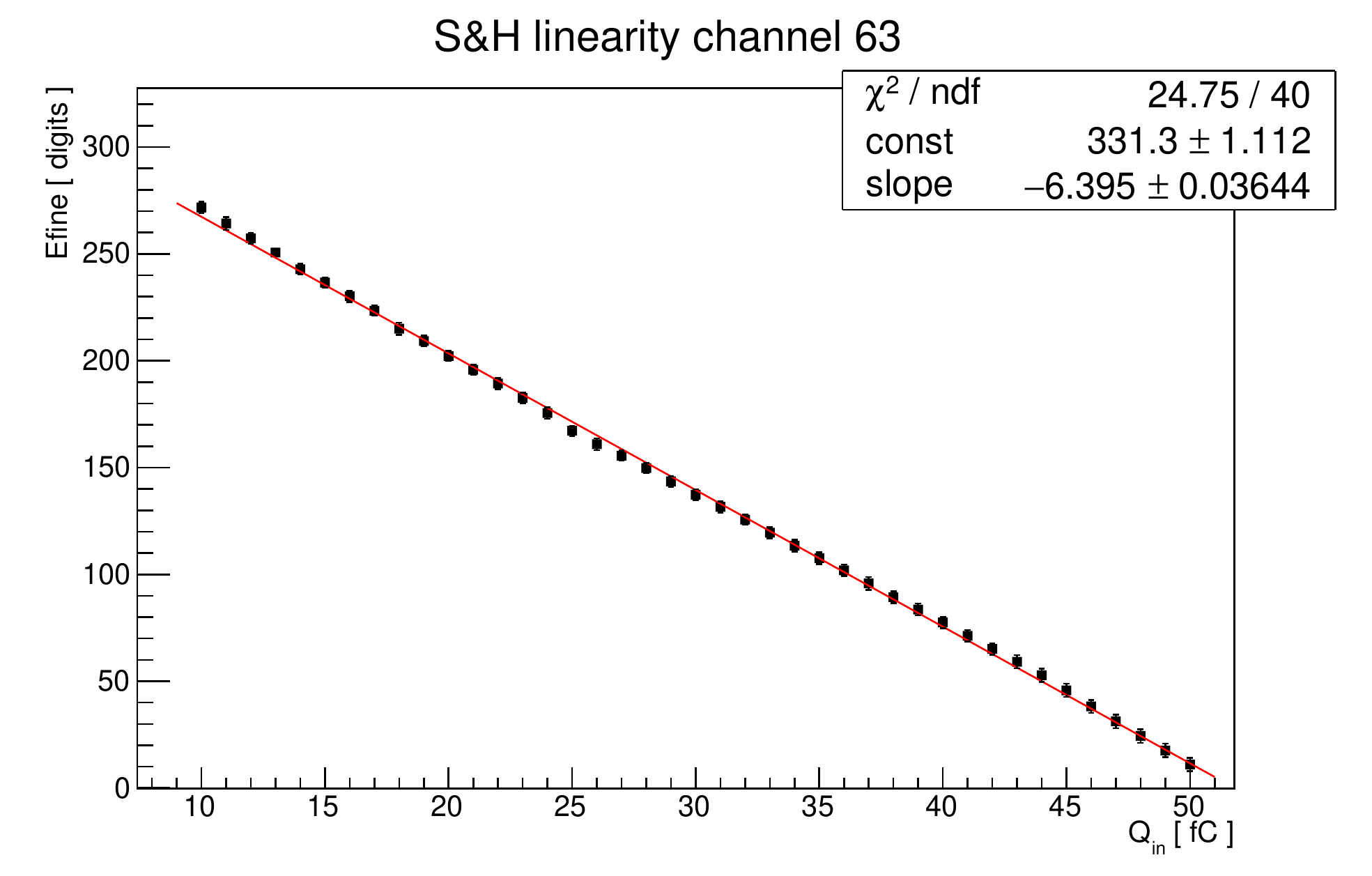}
\caption{\label{fig:charge} Charge measurement with time-over-threshold (\emph{left}); and sample-and-hold circuit (\emph{right}).}
\end{figure}

Figure \ref{fig:charge} shows the charge measurement results on a single test channel, using ToT and the S\&H circuit, and injecting a known input charge with an external programmable-amplitude pulse generator.
The results match the expected linearity (better than 0.2\%) of the sample and hold circuit up to an input charge of 50 fC.
The characterisation with an external test pulse generator is used for the calibration of the internal calibration signal, which is afterwards used for systematic testing.

\begin{figure}[htbp]
\centering
\includegraphics[width=.42\textwidth]{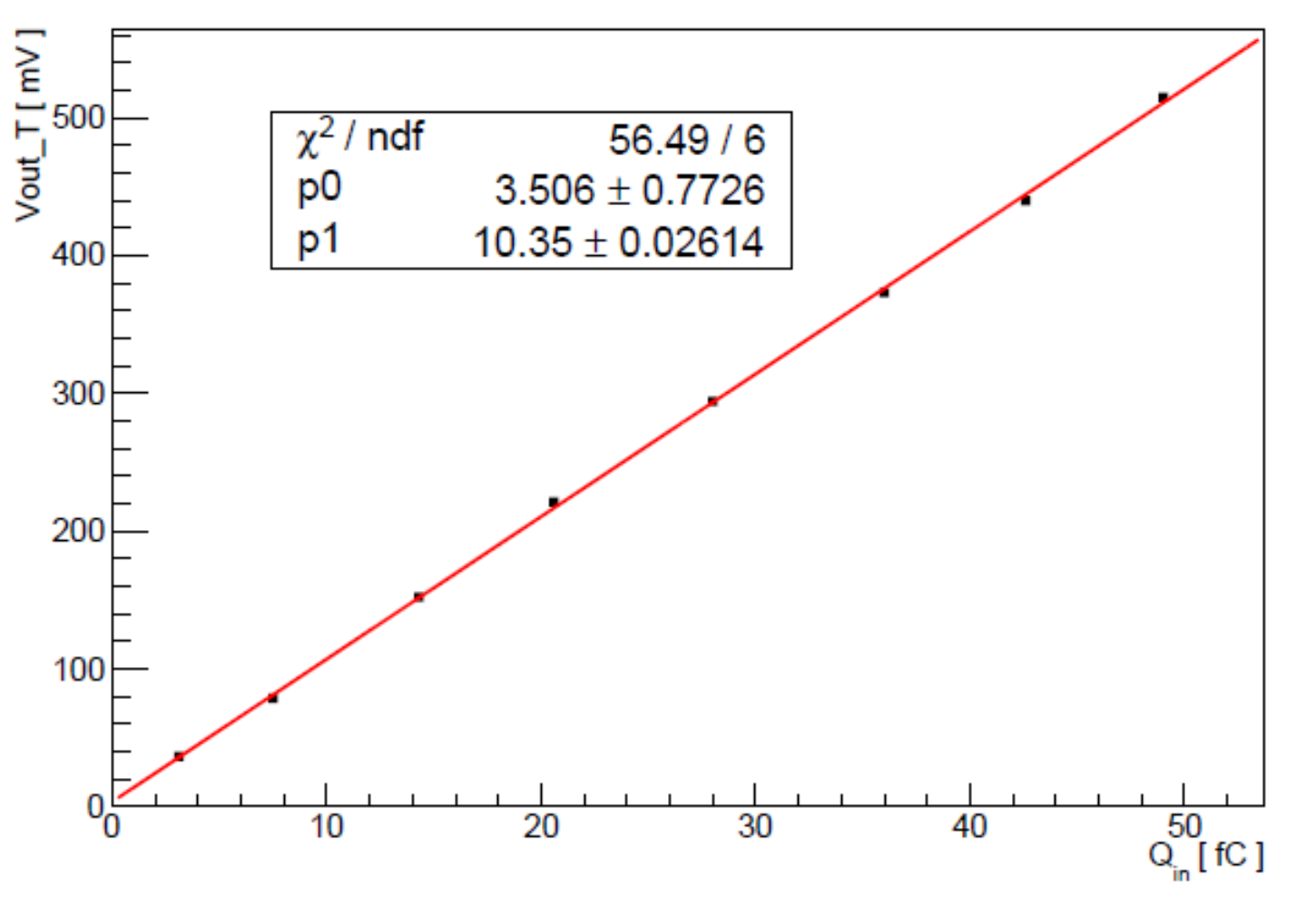}
\qquad
\includegraphics[width=.46\textwidth]{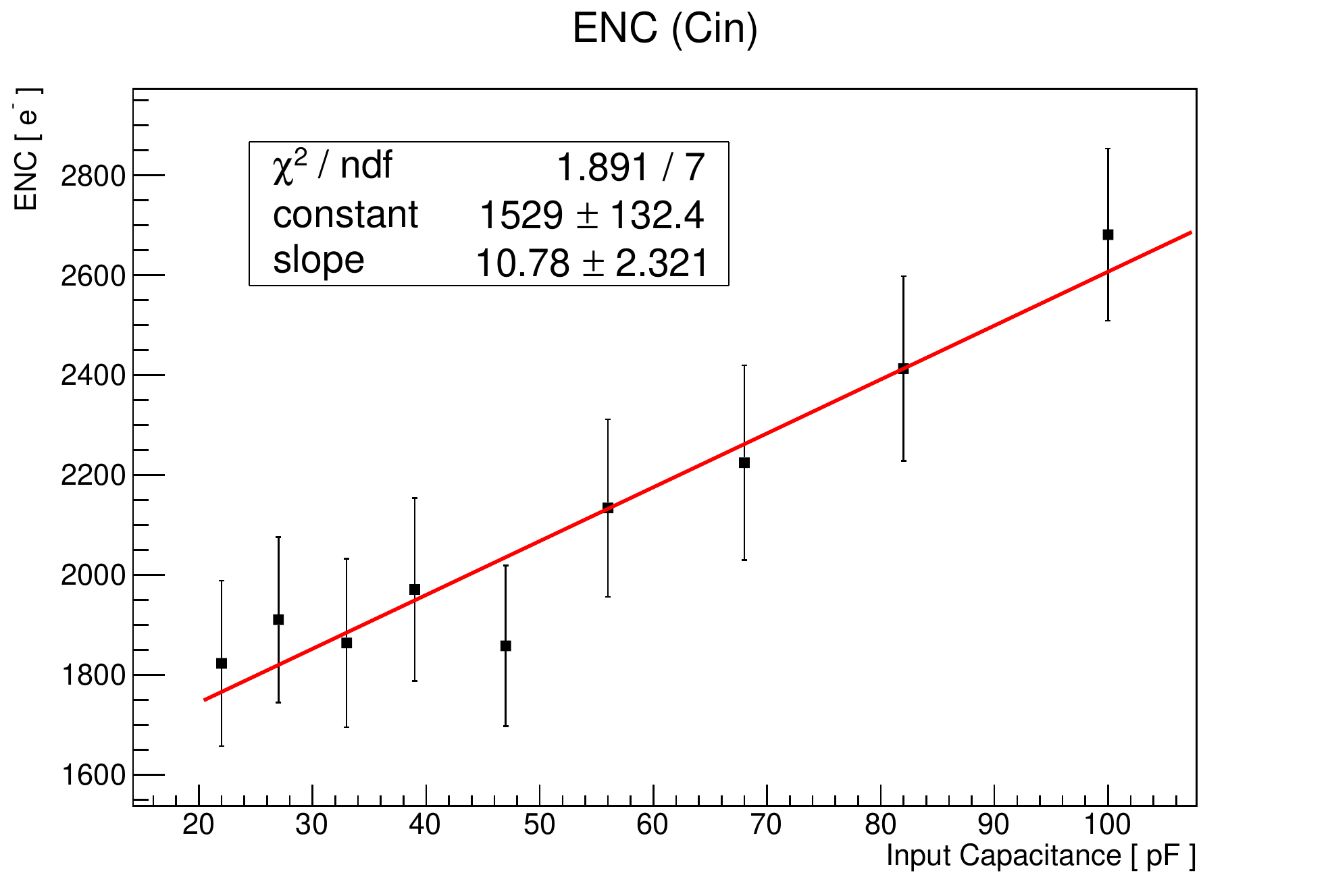}
\caption{\label{fig:gain} Gain measurement on a debug output (\emph{left}); and ENC of the slow shaper (\emph{right}) as a function of the input capacitance.}
\end{figure}

A set of probing points in one channel allows for a direct measurement of the pulse amplitude at the output of the fast shaper, as well as the LSB and offset of the discriminator threshold.
This feature thereby allows for a direct measurement of the amplifier gain in one of the channels.
The average gain of 10 mV/fC is in good agreement with post-layout simulation results (refer to figure \ref{fig:gain}, left plot), and the residual channel-to-channel dispersion is below 0.2 mV/fC r.m.s.
On the other hand, the equivalent noise charge (ENC) of 2600 electrons r.m.s. on the slow shaper output with an input capacitance of 100 pF is almost 30\% higher than expected.
However, the plateau of the noise characteristic at low input capacitance (refer to figure \ref{fig:gain}, right plot) seems to indicate that the measurement is affected by common-mode noise.
Front-end board and test setup grounding and shielding, as well as power-supply rejection ratio (PSRR) of the front-end blocks are under study in order to understand the root cause of this increase.

\section{Outlook and Conclusions}

The analogue readout of the CGEM detector for the BESIII inner tracker upgrade calls for an innovative custom detector electronics.
A dedicated 64-channel ASIC was designed to target the detector specifications.
Electrical characterisation of the first prototype was performed between October 2016 and May 2017, and detector test campaigns with radioactive sources and cosmic rays are ongoing since March 2017.
The volume production of the engineering version with minor revisions is planned for July 2017.
The assembly and qualification of the TIGER with the on-detector Front-End Boards is expected to start by Fall 2017.
System-level commissioning and instrumentation of the full CGEM detector will be started in February 2018, and the installation of the tracker is forecast for Summer 2018.

\acknowledgments

The research leading to these results has been performed within the BESIIICGEM Project, funded by European Commission in the call H2020-MSCA-RISE-2014.

\end{document}